\documentclass[prb,aps,reprint,longbibliography]{revtex4-2}

\usepackage[english]{babel}
\usepackage{graphicx}
\usepackage{color}
\usepackage{amsmath}
\usepackage{tabularx}
\usepackage[dvipsnames]{xcolor}

\begin{document}
\title{Evidence for correlated electron state from shot noise in ``bad” metal nanojunctions}

\author{Yiou Zhang}
\email{yiou.zhang@emory.edu}
\affiliation{Department of Physics, Emory University, Atlanta, Georgia 30322, USA}
\author{Chendi Xie}
\altaffiliation[Also at ]{Department of Physics and Astronomy, Clemson University, SC 29631, USA}
\affiliation{Department of Chemistry, Emory University, Atlanta, Georgia 30322, USA}
\author{John Bacsa}
\affiliation{Department of Chemistry, Emory University, Atlanta, Georgia 30322, USA}
\author{Yao Wang}
\affiliation{Department of Chemistry, Emory University, Atlanta, Georgia 30322, USA}
\author{Sergei Urazhdin}
\affiliation{Department of Physics, Emory University, Atlanta, Georgia 30322, USA}

\begin{abstract}
We report anomalously large shot noise in metallic and tunnel nanojunctions of $\beta$-tantalum, a metal characterized by a small negative temperature coefficient of resistivity inconsistent with both the Fermi liquid theory of Bloch electrons and single-particle localization. Fano factors are enhanced, and cluster around even multiples of the values expected for Fermi liquids, suggesting the possibility of pairwise electron correlations akin to local groups of Cooper pairs. The importance of spin for this state is suggested by the effects of magnetic impurities. Additional evidence for correlations is provided by the reduced density of states near the Fermi level revealed by point contact spectroscopy and first principles calculations. Our findings indicate that electron correlations may play an important role in a much larger class of materials than previously believed, opening new avenues for their studies and applications.

\end{abstract}

\maketitle

\section*{Introduction}

Exotic electronic states not described by the single-particle Fermi liquid (FL) picture can emerge close to the quantum critical points (QCPs) such as metal-insulator transition (MIT)~\cite{RevModPhys.60.585}. Based on the Ioffe-Regel criterion, electron localization is expected if the material resistivity $\rho$ exceeds a crossover value corresponding to the mean free path (mfp) shorter than the lattice constant, about $1.5\,\mu\Omega\cdot m$ in common transition metal alloys~\cite{Hussey2004}. However, a wide range of materials called ``bad" metals remain metallic in the Ioffe-Regel limit. ``Bad" metals often exhibit a small non-divergent negative temperature coefficient of resistivity (TCR), described as ``weak" or ``failed" insulator~\cite{zeng2021transport,zhang2022anomalous}. These behaviors are inconsistent with both single-electron localization and diffusive transport.

Among ``bad" metals are unconventional superconductors~\cite{ono2000metal,ando1995logarithmic,riggs2009magnetic} and other strongly correlated materials~\cite{sahoo2020pressure,lin2018structure,cao2020strange,li2020absence,zhang2018review}, suggesting that ``bad" metallicity may involve many-particle effects~\cite{zeng2021transport,hegg2021geometric,chowdhury2022sachdev}. A large magnetoresistance observed in some conventional superconducting (SC) ``bad" metal films close to the critical temperature $T_C$~\cite{kim2012superconductor,qin2006magnetically,breznay2017particle,breznay2017superconductor} was attributed to the correlated state termed the Bose liquid~\cite{christiansen2002evidence,phillips2003elusive,dubi2007nature,phillips2019free,yang2019intermediate, hegg2021geometric,zeng2021transport}. This state was hypothesized to host incoherent Cooper pairs~\cite{zhang2022anomalous} or a local pair condensate that lacks long-range phase coherence~\cite{PhysRevLett.95.077002}. However, there is presently no direct evidence for the Bose liquid state in ``bad" metals that are either non-SC or at temperatures far above their $T_C$. Since ``bad" metallicity often persists at room temperature, the existence of such a state may imply pairing mechanisms potentially relevant to high-temperature superconductivity. 

\begin{figure}
\centering
\includegraphics[width=1.0\columnwidth]{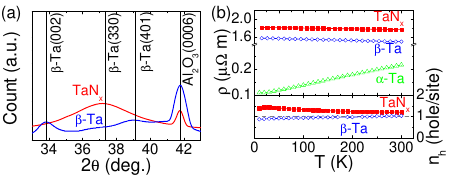}
\vspace{-5pt}
\caption{\label{fig:fig1} (a) XRD $\theta-2\theta$ scan on $\beta$-Ta(35) and TaN$_x$(35) on Al$_2$O$_3$ (0001) using Co $K_\alpha$ line. Tilting by $approx2^\circ$ off-Bragg was used to minimize substrate reflections. The curves are offset for clarity. The expected positions $2\theta=39.36^\circ$ and $48.84^\circ$ of the $\beta$-Ta(002) and Al$_2$O$_3$(0006) peaks are indicated. (b) Resistivity (top) and carrier concentration from the Hall data (bottom) for $15$~nm-thick $\beta$-Ta, TaN$_x$, and $\alpha$-Ta films, as labeled. Solid lines are fittings based on the e-e interaction-mediated hopping model~\cite{Szurek2025}.
}
\end{figure}

The $\beta$-allotrope of tantalum (Ta) is an archetypal elemental ``bad" metal and a ``failed" insulator [see Fig.~\ref{fig:fig1}(b)], which is widely used in spin-orbitronics as a spin current source due to its large spin Hall effect (SHE)~\cite{liu2012spin}. The $\alpha$-allotrope of Ta, a typical Fermi liquid, exhibits a significantly smaller SHE, suggesting that the mechanism responsible for ``bad" metallicity may be also important for spin Hall efficiency. 
Motivated by the observation that ``bad" metallicity of $\beta$-Ta cannot be explained by single-particle disorder, Refs.~\cite{szurek2024anomalous,Szurek2025} studied electron-electron (e-e) interaction effects by utilizing shot noise (SN) measurements in $\beta$-Ta nanowires and metallic junctions (MJs). SN is white noise caused by the discrete nature of charge carriers. For tunnel junctions (TJs), its current spectral density is $S_I=2FeI,$  where $I$ is current, $F=q/e$ is the Fano factor, $q$ is the charge of elementary excitations and $e$ is the electron charge, and thermal effects are neglected~\cite{ShotNoiseBible}. Since the Fano factor is dependent on the charge of elementary excitations, SN provides a sensitive measure of electron correlations. In particular, Fano factor doubling was demonstrated in TJs based on superconductors due to Cooper pairing~\cite{zhou2019electron,bastiaans2021direct}.

The properties of SN in non-SC TJs are expected to be independent of the tunnel barrier height and width as long as the transmission probabilities through all the conduction channels are substantially smaller than $1$. In contrast, MJs are characterized by the dominant transmission through almost open conduction channels, resulting in Ohmic transport. In this case, the SN properties are determined by the transport regime and electron-phonon interactions. Diffusive transport of charges $q$ in a short MJ is expected to give $F=q/3e$, while thermalization due to electron-electron (e-e) interaction is expected to result in the ``hot" electron regime where $F=\sqrt{3}/4$ is determined by the fluctuation-dissipation relation~\cite{HotElectrons}. SN suppression in ``strange" metal nanowires was attributed to a correlated electron liquid that lacks single-electron excitations~\cite{chen2023shot,zhang2024shot}. Similar behaviors were observed in $\beta$-Ta nanowires~\cite{szurek2024anomalous}, but were subsequently shown to result from electron-phonon interactions~\cite{Szurek2025}. Measurements of nanometer-scale $\beta$-Ta junctions revealed ``hot" electron regime in junction as short as $8$~nm, suggesting strong e-e interactions~\cite{Szurek2025}. However, the nature of many-electron states resulting from these strong interactions remains unknown. Notably, the superconducting transition temperature $T_C<1$~K in $\beta$-Ta~\cite{Read1965} is far below the temperature range where ``bad" metallicity and anomalous SN are observed. Thus, the underlying correlations are likely distinct from the conventional superconducting pairing, and may instead involve Mott-Hubbard mechanism~\cite{supp}.

Here, we present direct evidence for the correlated many-electron state in $\beta$-Ta distinct from incoherent Cooper pairing, by utilizing SN measurements in TJs and point-contact (PC) MJs. To the best of our knowledge, this work represents the first study of $\beta$-Ta based TJs. The effective length of PC MJs is significantly shorter than in the previously studied MJs, minimizing the electron thermalization effects that prevented characterization of carrier charge. Our results advance SN measurements as an efficient approach to studies of e-e interactions, and provide insight into the correlations underlying anomalous metallic states.

\section*{Methods} 
\begin{figure}
\centering
\includegraphics[width=1.0\columnwidth]{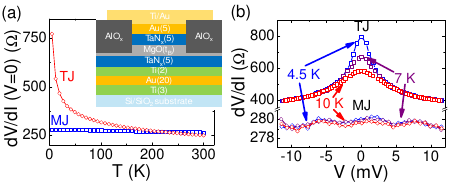}
\vspace{-5pt}
\caption{\label{fig:fig2} $R$ vs $T$ (a) and $V$ (b) for a TJ and MJ, as labeled. Inset in (a) is the cross-section schematic of the junctions.}
\end{figure}

The studied junctions were formed by Ta layers sandwiched between two current-carrying metallic Au electrodes [see inset in Fig.~\ref{fig:fig2}(a)]. It is established that pure $\beta$-Ta can be obtained by deposition directly on a sapphire substrate after a short Ar ion bombardment to activate its surface polar bonds~\cite{szurek2024anomalous}. We used this method to deposit the reference $\beta$-Ta films discussed below. In contrast, pure Ta grown on metallic buffer layers forms the $\alpha$ allotrope~\cite{szurek2024anomalous,Szurek2025}.  To overcome this issue, we developed a method to obtain a well-defined $\beta$ allotrope of Ta deposited on metallic layers. This was accomplished by sputtering $99.9\%$ pure Ta in the presence of a small partial pressure $P_N=4\times10^{-6}$~Torr of ultrapure nitrogen, in addition to the ultrahigh purity Ar processing gas at $P_{Ar}=4\times10^{-3}$~Torr, in ultrahigh-vacuum chamber with the base pressure of $4\times10^{-9}$~Torr. 

X-ray diffraction (XRD) for the test films deposited on a Ti(1) buffer layer [the number is thickness in nanometers] showed a single $\beta$-Ta (002) peak slightly shifted and reduced in amplitude compared to pure $\beta$-Ta, with no signatures of impurity or $\alpha$ phases [Fig.~\ref{fig:fig1}(a)]. Energy dispersive spectroscopy showed that a small amount of nitrogen became incorporated in the film, forming a nominal composition TaN$_x$ with $x=9\pm2\%$ (see Supplemental Materials, SM). However, the lack of noticeable XRD peak broadening shows that nitrogen was predominantly incorporated into the grain boundaries and/or as interstitial, without significantly changing the $\beta$-Ta structure. This is consistent with the slight shift of the XRD peak corresponding to the tensile lattice strain of $\approx0.3\%$.

Hall measurements gave carrier concentration of one hole per atom in $\beta$-Ta, and a slightly higher hole concentration in TaN$_x$, as expected from nitrogen doping. The Hall carrier concentrations show a small temperature dependence with opposite trends. To the best of our knowledge, there is no established general theory of the Hall effect in non-FLs, warranting further analysis of the mechanisms underpinning these differences.
The mean free path is somewhat reduced by nitrogen incorporation (see SM). The resistivity is slightly increased but its temperature dependence remains nearly the same as in pure $\beta$-Ta (Fig.~\ref{fig:fig1}(b)). Based on these results showing only minor effects of nitrogen doping, we conclude that the central findings for TaN$_x$ described below are also relevant to pure $\beta$-Ta.

The studied junctions were based on multilayers formed by two TaN$_x$ electrodes separated by an MgO tunnel barrier and sandwiched between conducting Au electrodes, inset in Fig.~\ref{fig:fig2}(a). We have also fabricated junctions based on AlO$_x$ tunnel barrier. They showed a very large flicker noise likely caused by the charge trapping in defects at the AlO$_x$/Ta interface, which prevented SN characterization in these junctions.

All the materials are DC-sputtered except for MgO, which is RF-sputtered. The multilayer films were vacuum-annealed at 300~$^\circ$C for $2$ hours and patterned into micron-scale junctions by e-beam lithography. MgO barrier thickness $t_B=2.9$~nm for TJs was chosen so that the junction resistance was in the range $R=0.2-2$~k$\Omega$, which provides optimal sensitivity of our noise measurement setup.

According to the established theories of tunneling  between metallic electrodes through an insulating barrier, the TJ resistance is expected to decrease with increasing bias and increasing $T$~\cite{Nordheim1928-gs,Simmons1963}. In contrast, transport through a pinhole in a TJ is Ohmic, i.e. its resistance is independent of bias and temperature.  Thus, the bias and the temperature dependence provide two distinct criteria for the junction characterization of junction quality, as illustrated in Fig.~\ref{fig:fig2} for a representative TJ. 

To independently confirm that the anomalous SN is not associated with specific junction chemistry, we also studied nanoscale MJs. The latter were naturally formed in some samples due to pinholes in the tunnel barrier, as was evidenced by the bias- and temperature-independent junction resistance distinct from the dependences observed for TJs. The naturally formed junctions were not sufficiently stable for SN measurements. Instead, the studied MJs were formed by applying voltage pulses until junction resistance dropped to $R\approx300 ~ \Omega$. Since the insulating barrier is known to break down in ``hot-spots" with the smallest local barrier thickness, such junctions form small pinholes whose resistance can be comparable to TJs. Since our noise measurement setup is designed for junction 
resistance $R=0.2-2$~k$\Omega$, we further optimized measurement sensitivity by using a slightly thicker $3.5$~nm MgO layer in these junctions. Before the voltage bias-induced breakdown, their resistance was in a few-k$\Omega$ range. Thus, shunting of the pinhole by conduction through the insulating layer was minimized. Because of the dominance of hot-spots, the
effective length of the junctions formed using this approach is likely significantly smaller than the nominal barrier thickness  $t_B=3.5$~nm, avoiding electron-phonon scattering effects that can compromised SN measurements in nanowires~\cite{szurek2024anomalous}. After the junction was formed, no further variations of the junction resistance was observed when additional voltage pulses of either polarity were applied. Thus, the possibility of memristive-like history dependence of junction properties~\cite{Chiu2012}, including the noise discussed below, can be eliminated. 

We emphasize that the junction resistance by itself does not allow one to distinguish TJs from MJs since it is dependent on the effective junction area, which can be much smaller for the MJs due to their pinhole conduction mechanism. For $R\approx300 ~ \Omega$, we estimate the effective junction size of $2-3$~nm. We confirmed the metallic conduction in our MJs by the lack of resistance dependence on $T$ or  $V$ (Fig.~\ref{fig:fig2}). This eliminates the possibility of memristive behaviors in either transport properties of noise producted by the junctions, since bias history dependence necessarily requires bias dependence.

To characterize the SN, voltage noise $S_{V}=(dV/dI)^2S_{I}$ was measured by the cross-correlation technique, as reported elsewhere~\cite{zhang2024shot} (see also SM). The noise spectra were analyzed to elucidate the contribution from the flicker noise (FN), which was identified by the $1/f$ spectrum, a quadratic increase with bias, and an increase with $T$~\cite{VANDERZIEL1950359}. Each data point was obtained by averaging multiple cross-correlated spectra in the noise frequency window where FN and EMI could be eliminated~\cite{supp}, with the acquisition time of $30-300$~sec per point depending on the frequency window. The calibration and measurement approaches were verified using Al/AlO$_x$/Al TJs, yielding Fano factor $F\,=\,1$ as expected for single-electron tunneling between Al electrodes~\cite{supp}.

\section*{Results}

\begin{figure}
\centering
\includegraphics[width=1.0\columnwidth]{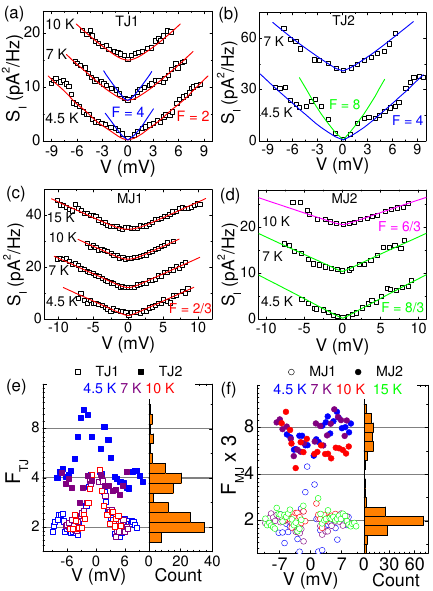}
\vspace{-5pt}
\caption{\label{fig:fig3} (a)-(d) Shot noise vs bias for two TJs (a),(b) and MJs (c),(d), at the labeled $T$. The noise data for different temperatures are offset for clarity. The curves are fittings with Eq.~(\ref{eq:noise_TJ}) for TJs and Eq.~(\ref{eq:noise_MJ}) for MJs. (e),(f) $F$ vs bias for TJs (e) and MJs (f), and histograms of the corresponding distributions.}
\end{figure}

Figures~\ref{fig:fig3}(a)-(d) show $S_I(V)$ for two TJs and two MJs. $9$ additional TJs and $7$ MJs show similar behaviors in the range of $T$ and $V$ limited by the onset of FN~\cite{supp}. For all the junctions, the noise increases approximately linearly with bias above a few mV, consistent with the expectations for SN. 

We used two complementary approaches to extract information about the charge carriers from these measurements.  In the first approach, the dependence of noise on bias was fitted using  a generalization of the forms expected for the single-electron Fermi liquids (FLs). The standard expression for noise produced by a FL TJs is~\cite{ShotNoiseBible,zhang2024shot}
$$
  S_I(V) = \frac{2eV}{R} \coth\frac{eV}{2k_BT},
$$
where $R=dV/dI$ is differential conductance and $k_B$ is the Boltzmann constant. The corresponding Fano factor is $F=1$. The noise power produced by tunneling of charges $q=F_{TJ}e$ is then
\begin{equation}\label{eq:noise_TJ}
	S_I(V)= \frac{2F_{TJ}eV}{R}\coth\frac{F_{TJ}eV}{2k_BT}.
\end{equation}
For single-electron diffusion in a FL MJ, both the random matrix approximation\cite{beenakker1992suppression} and semi-classical Boltzmann equation\cite{nagaev1992shot,nagaev1995influence} predict 
$$
	S_I(V)= \frac{2eV}{3R}\coth\frac{eV}{2k_BT} + \frac{8}{3}\frac{k_BT}{R},
$$
which gives the Fano factor $F=1/3$. An extension of this equation obtained by replacing $e$ with $F_{MJ}e$ is 
\begin{equation}\label{eq:noise_MJ}
  S_I(V)= \frac{2F_{MJ}eV}{3R}\coth\frac{F_{MJ}eV}{2k_BT} + \frac{8}{3} \frac{k_BT}{R}.
\end{equation}
These equations describe a crossover between a linear dependence dominated by SN at $V>\gg k_BT$, with the slope given by the Fano factor, and the thermal regime dominated by Johnson noise $S_v=4k_BTR$ at $V\gg k_BT$. An empirical form slightly different from Eq.~(\ref{eq:noise_MJ}) was proposed in Ref.~\cite{chen2023shot} for MJs, resulting in minor differences between the expected thermal broadening not essential for the analysis below.

The Fano factors extracted from the fittings for all the studied TJs and MJs with with Eqs.(\ref{eq:noise_TJ}), (\ref{eq:noise_MJ}), respectively, are anomalously large, and cluster around even multiples (2, 4, 6, and 8) of the values expected for FLs (Fig.~\ref{fig:fig3}(a)-(d) and SM). We do not observe any noticeable correlation between junction resistance and variations in $F$ among different junctions. The Fano factors in the TJs are enhanced at small $V$, which is reminiscent of the increase due to Cooper pairing in SC TJs at bias below the SC gap~\cite{lefloch2003doubled,bastiaans2021direct}. The uncertainties in our fitting of the limited data set leave room to alternative interpretations of Fano factor enhancement at small bias. Nevertheless, these features are symmetric with respect to the bias polarity, confirming that they are not random noise variations. We also note that the possible artifacts due to FN are minimized at small bias, and cannot account for the observed Fano factor enhancement. 

Importantly, the enhanced Fano factors are almost temperature-independent in MJs. For TJs, they show a decrease at small bias but remain enhanced. These observations indicate that the correlations that underpin the enhanced SN originate from a robust interaction mechanism not directly related to superconducting ordering observed in beta-Ta only below $1$~K~\cite{PhysRev.119.1578}.

To further test our observation that $F$ may cluster around even multiples of values expected for FLs, we used an alternative approach to data analysis based on the observation that the magnitude of noise at a single bias point $V\gg k_BT$ can characterize the carrier charge. We define the Fano factor for individual data points as $F_{TJ} = S_IR/2eV$ for TJs, and $F_{MJ} = (S_IR-8k_BT/3)/2eV$ for MJs. The former expression neglects thermal effects whose relative contribution is larger in MJs due to smaller F, and is accounted for in the latter expression. The calculated distribution of $F$ is quasi-continuous due to transitions between different regimes and thermal broadening at small $V$. Nevertheless, both $F_{TJ}$ and $3F_{MJ}$ cluster around even values, Figs.~\ref{fig:fig3}(e),(f). We note that this result may be influenced by artifacts due to uncertainties stemming from the limited noise statistics and a possible non-negligible contribution from FN. Point-to point variation in Figs.~\ref{fig:fig3}(a)-(d) provides an estimate of random noise comparable to the symbol size. However, the potential systematic uncertainty from FN is more difficult to estimate. 

\begin{figure}
\centering
\includegraphics[width=1.0\columnwidth]{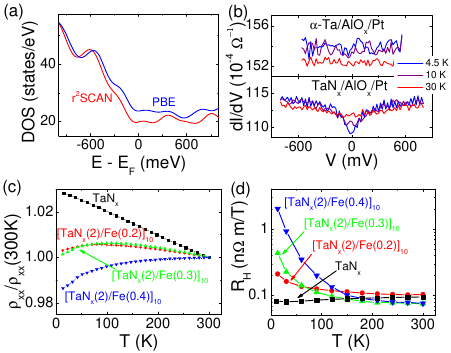}
\vspace{-5pt}
\caption{\label{fig:fig4}  (a) DOS of $\beta$-Ta calculated using the r$^2$SCAN mGGA and PBE functionals. (b) Conductance spectra of TaN$_x$ and $\alpha$-Ta PCs. (c) Resistivity and (d) Hall coefficient of Ta/Fe multilayers vs $T$. }
\end{figure}

The results of noise measurements indicate that charge current in the studied TJs and MJs may be carried by electron groups rather than single electrons, suggesting a correlated electron state. We now present additional theoretical and experimental evidence supporting this possibility. Because of the lattice complexity and multi-orbital electronic structure of $\beta$-Ta, many-body simulations that could verify whether anomalous SN in $\beta$-Ta may be caused by correlations are impractical. Instead, we performed comparative first principles band structure calculations based on two different approximations for the exchange-correlation functional: the Perdew-Burke-Ernzerhof (PBE)  and r$^2$ strongly constrained and appropriately normed (r$^2$SCAN) meta-generalized gradient approximation (mGGA)~\cite{furness2020accurate}. Neither the PBE nor the r$^2$SCAN approximations capture many-particle effects~\cite{huang2016much}, but the latter better accounts for correlations~\cite{Furness2020,kingsbury2022performance}. Thus, their differences can provide insight into correlations even in single-particle approximation.

 The calculations utilized the \textsc{Quantum ESPRESSO} package~\cite{giannozzi2017advanced}, which is based on the plane-wave pseudopotential method. The PBE calculations were based on the exchange-correlation functional with SG-15 optimized norm-conserving Vanderbilt (ONCV) pseudopotentials~\cite{perdew1996generalized,hamann2013optimized}. Structural relaxation was performed using the r$^2$SCAN~\cite{furness2020accurate} functional with a semiempirical Grimme’s DFT-D2 van der Waals correction~\cite{grimme2006semiempirical}. The r$^2$SCAN electronic structure calculations were performed using the same r$^2$SCAN functional with the relaxed structures. A cutoff energy of $200$~Rydberg was used to expand the wavefunctions, and a $\Gamma$-centered Monkhorst-Pack sampling ~\cite{monkhorst1976special} of $4\times4\times8$ was adopted for the integral over the Brillouin zone. The convergence criterion of the corresponding electronic self-consistent field was set to 10$^{-10}$ Rydberg.

 The density of states (DOS) calculated using  r$^2$SCAN is reduced near the Fermi energy compared to the PBE functional, Fig.~\ref{fig:fig4}(a). Comparing the calculated band structures (see SM), this is associated with the overall shift of the bands away from the Fermi surface, consistent with the effects of correlations. We emphasize that the band picture is inconsistent with the ``failed" insulator properties of $\beta$-Ta, so these calculations provide only single-particle signatures of the many-electron correlation effects, and are not capable of capturing their magnitude.

 To investigate the possibility of DOS suppression in $\beta$-Ta, we performed additional point-contact (PC) spectroscopy. The PCs were fabricated from multilayers similar to those in SN  measurements, except the MgO barrier was replaced by sputtered AlO$_x$ ($t_B~=~3.5~nm$) and annealing was not performed, resulting in metallic point-contacts (PCs) through pinholes in AlO$_x$. Additionally, the TaN$_x$ layer on top of AlO$_x$ was replaced by Pt, so that the PC spectra were determined entirely by the electronic structure of the bottom electrode, either TaN$_x$ or $\alpha$-Ta. The latter was grown by sputtering Ta in pure Ar, as discussed above. 

The differential conductance of a PC based on $\alpha$-Ta is independent of bias, as expected for weakly energy-dependent DOS in the FL, Fig.~\ref{fig:fig4}(b). In contrast, the conductance of PC based on $\beta$-Ta decreases at small bias, which becomes more pronounced at lower $T$. The variation occurs on a significantly larger bias scale than in MgO (Fig.~\ref{fig:fig2}(b)) and Al/AlO$_x$/Al TJs (see SM), and conductance is significantly higher than in TJs. This cannot be explained by tunneling effects but is consistent with reduced DOS near the Fermi level. The symmetric PC spectra do not reflect the asymmetry of the calculated band structure [Fig.~\ref{fig:fig4}(a)], confirming that the single-electron band approximation is inadequate.
The symmetric bias dependence also indicates that the junction resistance is not influenced by the Schottky effect.

If electron correlations are spin-dependent, they are expected to be suppressed by magnetic field. In the qualitative picture discussed below, these correlations involve electrons quasi-localized on the neighboring sites. The required fields are then of the order $B\sim \Phi_0/a^2\sim 10^4$~T, too large for direct testing of field effects. Here, where $\Phi_0$ is the flux quantum and $a=0.3$~nm is interatomic spacing. On the other hand, effective exchange fields in magnetic materials are of the same order of magnitude. 
To test their effects on $\beta$-Ta, we introduced proximity-induced effective exchange field via magnetic impurities, in multilayers with the structure [TaN$_x$(2)Fe($t_{Fe}$)]$_{10}$AlO$_x$(3) where $t_{Fe}$ ranged between 0.2~nm and 0.4~nm. The films were annealed in vacuum at $300^\circ$ for 2 hours to promote the formation of magnetic nanoclusters. In contrast to TaN$_x$, these films exhibit a positive TCR at cryogenic $T$, with the crossover temperature to negative TCR that increases with $t_{Fe}$, Fig.~\ref{fig:fig4}(c). This cannot be explained by the parallel conductance though Fe, in which case the opposite temperature dependence would be observed due to the saturation of Fe resistivity at cryogenic $T$. The onset of positive TCR coincides with the enhancement of Hall coefficient (Fig.~\ref{fig:fig4}(d)), as expected from the anomalous Hall contribution due to the onset of magnetism in Fe clusters. We conclude that ``bad" metallicity is suppressed by magnetism, supporting its origin from Mott singlet-like correlations. The large magnitude of effective exchange field needed to suppress this state confirms that the correlations are short-ranged.

\section*{Discussion}

Our central result is the observation of enhanced SN in TaN$_x$ nanojunctions, suggesting a correlated state. First, we discuss potential alternatives. Enhanced noise cannot be an artifact due to flicker noise (FN), as we do not observe any correlation between $F$ and FN among different samples, and the largest enhancement of $F$ in TJs is at small bias where FN is negligible. Cooper pair tunneling in SC junctions results in doubled $F$, and Andreev reflections due to SC gap can further enhance SN~\cite{PhysRevLett.83.2050}. However, this mechanism is not relevant to gapless non-SC $\beta$-Ta.  

SN enhancement in TJs can be caused by intermittent blocking of tunneling through a localized state in the barrier, due to interaction with another localized state~\cite{safonov2003enhanced,garzon2007enhanced,zhang2019spin}. However, analysis shows that this mechanism generally results in asymmetric dependence on bias direction, and even in special cases where it is symmetric $F$ should continuously evolve with bias, inconsistent with our results (see SM for details). 

We now propose a qualitative microscopic model for our observations. Each atom in $\beta$-Ta is coordinated by $20$ nearest or next-nearest neighbors within a narrow distance range $d=0.28-0.284$~nm~\cite{Magnuson2019}, resulting in a low-symmetry but almost isotropic crystal field. Large spin-orbit coupling (SOC) splits the $5d$ levels into the $J=5/2$ sextuplet and $J=3/2$ quadruplet. The latter is occupied by the three valence electrons, consistent with the measured charge density of $1$ hole per atom.

To analyze the electronic properties of $\beta$-Ta, we calculate the Koster-Slater amplitudes for hopping among the $J=3/2$ states. These amplitudes are determined by two matrix elements, $V_1=V_{dd\sigma}+V_{dd\pi}-2V_{dd\delta}$, $V_2=2(V_{dd\pi}+4V_{dd\delta})$, where $V_{dd\sigma}=6V_{dd\delta}=-3V_{dd\pi}/2$ in the atomic sphere approximation~\cite{andersen1978electronic,jenke2021tight} (see SM). Both $V_1$ and $V_2$ vanish, so only hybridization beyond this approximation contributes to hopping. Hopping suppression is evident from the small bandwidths $2t\lesssim 0.5$~eV in the first-principles calculations (see SM), significantly smaller than the Mott-Hund's energy $U=2-3$~eV~\cite{PhysRevLett.125.096403}.

In single-orbital systems, the condition $t\ll U$ implies that charges become localized at half-filling, or form a strongly correlated state at sufficient deviations from half-filling. The effects of large interaction in multi-orbital systems are less explored~\cite{PhysRevB.83.205112}. Our Mott-Hubbard analysis shows that in this case, dephasing of single-particle wavefunction due to electron interactions also leads to localization, which for multi-orbital systems does not result in MIT due to the absence of Pauli blocking (see SM for details). The temperature-dependent resistivity of $\beta$-Ta and TaN$_x$ can be accurately fitted by the parallel-resistor formula~\cite{Hussey2004} based on a combination of the usual Mott's phonon-mediated quasi-localized charge hopping and e-e interaction-assisted hopping, as shown by solid lines in Fig.~\ref{fig:fig1}](b). The phonon contribution to hopping decreases with decreasing $T$ while the e-e assisted contribution is temperature independent, resulting a large overall resistivity and small non-divergent negative TCR in the regime dominated by e-e interactions. This interpretation accounts for the ``bad" metallicity and the ``failed" insulator behaviors.

The Mott-like localization can also account for the observed enhanced SN, as follows. Localization results in a metallic state where both hole localization and hopping are determined by interaction with other holes. Consequently, bunching effects reminiscent of charge blockage due to the localized charge trapping may be expected. In contrast to single-charge trapping on defects manifested by large and asymmetric bias dependence due to the sigle-particle energy conservation~\cite{garzon2007enhanced}, interaction-driven hopping involves energy transfer among single-particle states explaining the lack of such features for $\beta$-Ta.

Next, we show that the proposed picture explains our tentative observation of Fano factor clustering around even multiples of the FL values, as well as the observed competition between ``bad" metallicity and ferromagnetism. The latter observation suggests spin-dependent pairing incompatible with magnetism. The kinetic energy of holes quasi-localized on the neighboring sites is reduced by maximizing virtual hopping, which for a single orbital results in the Mott singlet valence bond (VB) correlations analogous to Cooper pairs, but mediated instead by Coulomb interactions and the Pauli principle. This scenario is not directly applicable to the multi-orbital $J=3/2$ quadruplet. However, the low-symmetry crystal field splits the quadruplet into two Kramers pairs and reduces SOC spin-mixing due to the hybridization with the $J=5/2$ states. Consequently, single orbital-like pseudo-spin SOC VB correlations due to Mott-Hund's interactions are expected  (see SM for details), explaining the observed competition with ferromagnetism. A similar mechanism was proposed for SrIrO$_3$, which also exhibits negative TCR at cryogenic $T$~\cite{zhang2024shot}.

\begin{figure}
\centering
\includegraphics[width=1.0\columnwidth]{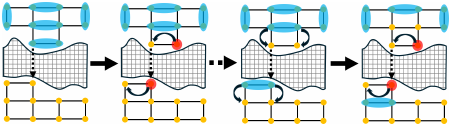}
\vspace{-5pt}
\caption{\label{fig:fig5} Correlated tunneling of holes through a hot-spot. The blue ovals are VB-like paired states, the red circles are unpaired holes. The dashed arrows indicate tunneling, curved arrows indicate hopping in the electrodes.}
\end{figure}

Because of the structural complexity and multi-orbital effects, quantitative analysis of tunneling involving many-electron states in $\beta$-Ta is outside the capabilities of modern computational techniques and will not be attempted here. Qualitatively, we argue that VB-like correlations favor bunched tunneling of even-numbered groups of holes, via the mechanism analogous to charge blockade by localized states in the tunnel barrier but involving quasi-localized holes at the interface. In contrast to Bloch states, tunneling of quasi-localized particles is likely dominated by atomic-scale hot-spots. If a hole tunnels out of such a hot-spot, the neighboring hole paired to this site by a VB (and which has a large amplitude on this site) is in a higher-energy state due to the lost correlation energy, resulting in an increased probability of its tunneling, as illustrated in Fig.~\ref{fig:fig5}. As both holes tunnel, other holes are attracted into the hot-spot by Coulomb effects leading to a cascade of tunneling. This proposed effect is similar to trapped charge-induced blockage, where the first tunneling pair plays the role of the trapped charge blocking the tunneling of other hole pairs. 

The characteristic number of bunched holes is likely determined by the local environment in the disordered metal, resulting in different values of $F$ among the junctions. However, pairing itself is a robust two-site effect, providing a tentative  explanation for the large observed variations of $F$ among diferent junctions, which nevertheless cluster around even values. 

Because of the competition between disorder and correlations, dynamical disorder should result in reduced bunching, consistent with decreasing $F$ when $T$ is increased (Fig.~\ref{fig:fig2}). We attribute the reduction of the Fano factor in TJs with increased bias to a similar mechanism of interaction of localized holes with bias-driven hopping of other holes through the same sites (see SM for details). 
These mechanisms should be strongly dependent on the atomic environment, which is expected to be manifested by large local variations in measurements by scanning SN microscopy~\cite{bastiaans2019imaging}.

\section*{Summary}
We presented evidence for a correlated state of quasi-localized electrons manifested by anomalous SN in junctions of a disordered ``bad" metal $\beta$-Ta. ``Bad" metals include a variety of functional materials, among them superconductors with low kinetic inductance and relatively high $T_c$ desirable for photon detectors and quantum information science applications~\cite{osofsky2001new,wakasugi1997superconductivity,mandal2020destruction,chauhan2022anomalously}. Importantly, the symmetry of VB-like singlets due to Mott-Hund's interactions identified in our analysis is the same as that of phonon-mediated Cooper pairs. Thus, superconductivity can be enhanced in these materials by unconventional contributions~\cite{doi:10.1021/jacs.4c06836}, warranting further studies of correlations in the non-SC state. 

Our results have significant implications for spin-orbitronics. Contrary to the common interpretations of SHE in $\beta$-Ta in terms of its band structure, the single-particle band picture is not applicable to the correlated quasi-localized holes. Remarkably, $\alpha$-Ta, which is a normal FL, exhibits almost negligible SHE. The relation between ``bad" metallicity and spin-orbitronic efficiency is supported by a giant (order of magnitude) enhancement of SHE in Pt alloys approaching the Ioffe-Regel limit~\cite{Shashank2025,liu2025enhancingzspingeneration}. Our analysis indicates that small crystal field effects experienced by quasi-localized electrons enable unquenched atomic orbital moments and consequently a large orbital Hall effect, and thus SHE. The effects of electron correlations on SHE can facilitate efficient correlated orbitronic and spin-orbitronic materials.

\section*{Data Availability}
The data that support the findings of this study are available on Zenodo at doi.org/10.5281/zenodo.17268050, and from the corresponding author upon reasonable request.

\section*{Acknowledgments}
We thank Connie Roth and James Merrill for assistance with spectroscopic ellipsometry. Experiments (Y.Z. and S.U.) were supported by the NSF award ECCS-2448290 and in part by the SEED award from the Research Corporation for Science Advancement. Simulations (C.X. and Y.W.) were supported by the U.S. DOE Office of Science BES Early Career Award No.~DE-SC0024524. The simulations used resources of the National Energy Research Scientific Computing Center, a U.S. DOE Office of Science User Facility at Lawrence Berkeley National Laboratory operated under Contract No.~DE-AC02-05CH11231.



\bibliography{TaTJ}


\end{document}